\documentclass[showpacs, amsmath, amssymb, pre, aps, reprint,
  floatfix]{revtex4-1}

\usepackage{graphicx}
\usepackage{braket}
\usepackage{natbib}

\begin{document}

\title{Universal behavior in finite 2D kinetic ferromagnets}

\author{J.\ Denholm}
\email{j.denholm@strath.ac.uk}
\author{B.\ Hourahine}
\email{benjamin.hourahine@strath.ac.uk}

\affiliation{SUPA, Department of Physics, University of Strathclyde, Glasgow, G4
  0NG, Scotland, UK}

\date{\today}

\begin{abstract}
We study the time evolution of the two-dimensional kinetic Ising model in finite
systems with a non-conserved order parameter, considering nearest-neighbour
interactions on the square lattice with periodic and open boundary
conditions. Universal data collapse in spin product correlation functions is
observed which, when expressed in rescaled units, is valid across the entire
time evolution of the system at all length scales, not just within the time
regime usually considered in the dynamical scaling hypothesis. Consequently,
beyond rapidly decaying finite size effects, the evolution of correlations in
small finite systems parallels arbitrarily larger cases, even at large fractions
of the size of these finite systems.
\end{abstract}


\maketitle

\section{Introduction}

\begin{figure}[h!]
  \includegraphics[width=\columnwidth]{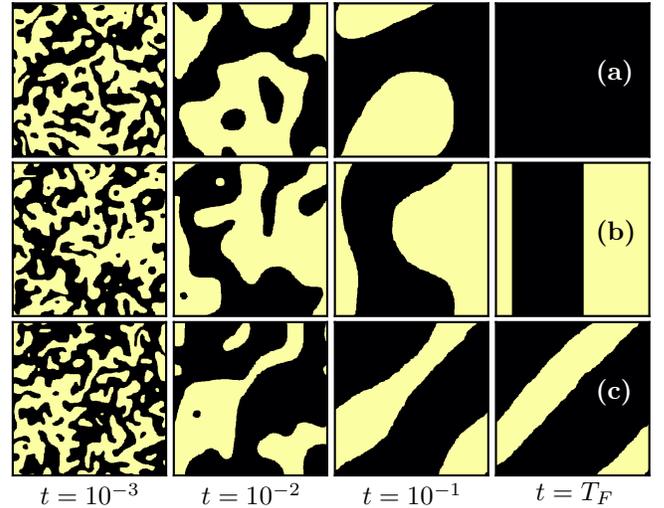}
  \caption{\label{fig:quenching_types} Snapshots of zero--temperature coarsening
    from random initial conditions on square lattices of $L=2^{10}$ for
    realizations leading to (a) stable, (b) metastable and (c) long--lived
    diagonal stripe configurations. The latter snapshots in (a) and (b) are the
    final states, reached at $T_{F}\approx0.3$ and $T_{F}=0.8$ respectively. The
    latter in (c) was halted early at $T_{F}=10.0$. Times are in arbitrary
    units.}
\end{figure}

Coarsening and phase separation in cooling systems have been studied extensively
and are generally understood in terms of curvature driven evolution at domain
interfaces~\cite{Lifshitz1962, Allen1979, Ohta1982}. The two--dimensional
kinetic Ising model is one such system. When quenched suddenly to below its
critical temperature, an elegant and beautiful phase separation process ensues:
magnetic domains nucleate and coarsen until a spanning domain structure forms
(see Fig.~\ref{fig:quenching_types}). Coarsening features found in the Ising
model have been observed in systems as diverse as binary Bose
gases~\cite{Tojo2010, Kawaguchi2010, De2014, Hofmann2014,
  Shitara2017,Takeuchi2018}, bacteria colony models~\cite{McNally2017} and
optical parametric oscillator systems~\cite{Oppo2001}.

The final state of quenching the nearest-neighbor Ising model to
zero-temperature depends on the spatial dimensionality of the system. In
one--dimension the ground state is always reached~\cite{Skorupa2012}, while in
three--dimensions the final states are a host of topologically complex
configurations that are forever trapped at constant energy in a local
minima~\cite{Olejarz2011a, Olejarz2011b, Spirin2002}. In two--dimensions, one
finds not only the ground state, but also ``frozen'' on-axis stripe states, or
long--lived off-axis stripe evolutions~\cite{Spirin2002, Spirin2001, Barros2009,
  Olejarz2012}. For nearest-neighbor interactions, the on-axis stripes are
infinitely long--lived and the off--axis stripes eventually decay to homogeneity
on a timescale of $\mathcal{O}(L^{3.5})$~\cite{Spirin2001}.

In two-dimensions, the probability of observing each topologically distinct
behaviour (Fig.~\ref{fig:quenching_types}) appears exactly equal to the
equivalent spanning probability in continuum percolation~\cite{Barros2009,
  Olejarz2012}, and has been further examined on various lattice
geometries~\cite{Cugliandolo2016, Blanchard2013}. The ability to identify the
``fate'' of a quench early in its evolution is rooted in the connection to
percolation, and gives an understanding of how the final state is
reached~\cite{Spirin2001, Barros2009, Olejarz2012}. Metastability has been
further explored in two--dimensions~\cite{Olejarz2013, Yu2017, Mullick2017}. The
scaling behavior of the ``fate sealing'' time has also been
studied~\cite{Blanchard2014a} and the connection with percolation investigated
with dynamics other than those of Glauber~\cite{Godreche_2018}.

The coarsening dynamics exhibited in the kinetic Ising model is generally well
understood through the phenomenology of the dynamical scaling hypothesis. This
states that the time evolution of the system is governed by a \emph{single}
relevant process, the growth of a characteristic length scale~\cite{Bray1993},
generally taken to be the typical domain size. The dynamical scaling hypothesis
is usually concerned with the infinite lattice case, thus (in its simplest form)
only applies in finite systems when the typical domain size is much less than
the system length.

Despite its success, there are few systems where the validity of this hypothesis
has been analytically proven~\cite{Arenzon2007}. Interestingly, it has been
shown that in the early time regime of the Ising evolution there is another
growing length scale, signifying the approach to critical
percolation~\cite{Blanchard2014a, Cugliandolo2016, Corberi2017}. Scaling laws
associated with different time correlation functions in two and three spatial
dimensions have recently been explored~\cite{Vadakkayil2019}.

In this work we examine the phase ordering kinetics of the non-conserved
two-dimensional Ising ferromagnet that is quenched to zero-temperature from
random initial conditions. We introduce the model and simulation method in
Sec.~\ref{sec:methods}. In Sec.~\ref{sec:results} we present the distribution of
times taken to reach a final state, as well as the time evolution of the energy
for each of the topologically distinct behaviors studied. Finally, we show our
main result, the data collapse in two-point same-time correlation functions
(also in Sec.~\ref{sec:results}). This applies throughout the entire evolution
of the system and not just within the usual dynamical scaling regime.


\section{Model and methods}
\label{sec:methods}

The cooperative nature of ferromagnets is such that their kinetic evolution can
be captured with short range interactions~\cite{Olejarz2012}. We accordingly
consider nearest-neighbor interactions on square lattices of length $L$. Spins
are denoted by $\mathcal{S}_{i} = \pm 1$ and can be viewed as a binary mixture
of phases with a non-conserved order parameter (or a non-conserved scalar
field). The total energy of the system is given by the Hamiltonian
\begin{eqnarray}
  \mathcal{H} =-\mathcal{J}\sum\limits_{ij}\mathcal{S}_{i}\mathcal{S}_{j},
\end{eqnarray}
where $\mathcal{J}>0$ is a ferromagnetic coupling constant and $j$ indexes the
nearest neighbors of each spin $\mathcal{S}_{i}$.

In accordance with zero-temperature Glauber dynamics, single spin flip events
that would decrease or conserve the system's energy assigned kinetic rates of
$1$ or $0.5$ respectively~\cite{Glauber1963}. Events that would incur an energy
cost are forbidden. We sample events using the $n$--fold method~\cite{bortz1975,
  Landau2009} and advance the time for each flip by $(\sum R_{k})^{-1}$, where
$R_{k}$ is the kinetic rate associated with each of the $k$ possible
events. This is the mean of the $n$-fold time update~\cite{bortz1975,
  Landau2009}.

For simplicity we present data for periodic boundary conditions with the mean
$n$-fold time update only. But we also find qualitatively similar behavior for
open boundaries and alternative time updates of either $n$-fold itself or
advancing time by the inverse of the number of active sites before each flip.

We obtain converged expectations from ensembles of $10^{4}$ quench trajectories
on systems of lengths $L = 2^{n}$ for $5 \le n \le 10$. In order to mimic
infinite temperature initial conditions, we initialize each quench by randomly
ordering microstates of zero net magnetization. The quenches are categorized by
the topologically distinct nature of their behaviours as cases reaching the
ground state, on--axis stripes or off--axis winding configurations.

We identify the long--lived configurations by searching for off--axis winding
domains at time $t\approx0.4L^{2}$, as by this time they are fully formed and
easy to identify (see Fig.\ref{fig:quenching_types}).

From each realization we extract two-point same-time correlation functions from
spins separated by on--axis distances $r=|\vec{r}_{i} + \vec{r}_{j}|$, and
obtain expectations of the form
\begin{equation}
  \mathcal{C}(r, t) =
  \braket{\mathcal{S}_{i}(\vec{r}_{i})\mathcal{S}_{j}(\vec{r}_{i} +
    \vec{r}_{j})},
  \label{eqn:correlation}
\end{equation}
where the averaging is over each spin in the system, the four $\pi / 2$
rotations of $\vec{r}_{i} + \vec{r}_{j}$ and across the ensemble of
simulations. Due to the irregular time step and duration of each quench, we
normalize the time for each realization to reach its final state and linearly
interpolate the correlations~\cite{scipy} to provide a regular time scale for
averaging. Realizations that evolve through the off--axis winding configurations
are halted early at $t=2L^{2}$ due to their long--lived nature. Correlations are
sampled for all $n$ such that $1 \le r \le L/2$ with $r = 2 ^{n}$. We specify
time values at which to pause each simulation and sample the correlations and
other observables.


\section{Results}
\label{sec:results}

\subsection{Quench time distribution}

There are two timescales associated with the zero--temperature coarsening of the
kinetic Ising model with periodic boundary conditions. The first is conventional
$\mathcal{O}(L^{2})$ coarsening, and is associated with realizations that reach
either a ground or on--axis stripe state~\cite{Spirin2001}. The second is
$\mathcal{O}(L^{3.5})$, and is associated with the slow decay of the off--axis
stripe configurations~\cite{Spirin2001}.

\begin{figure}[h!]
  \includegraphics[width=\columnwidth]{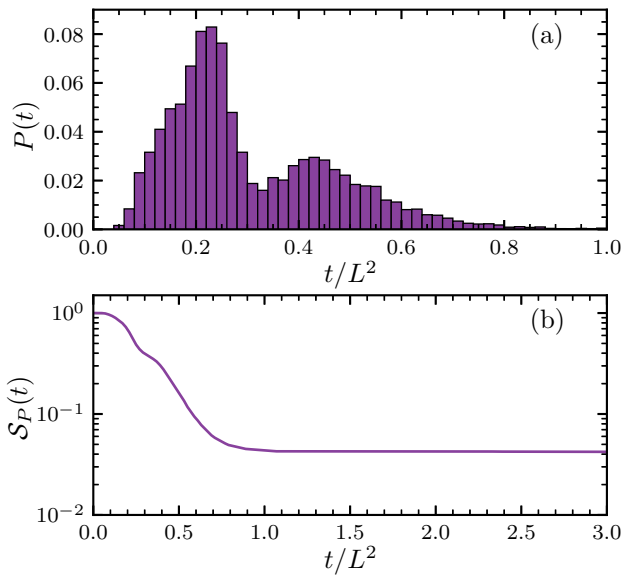}
  \caption{\label{fig:time_distributions} (Color online) (a) quench time
    distribution $P(t)$ and (b) the survival probability
    $\mathcal{S}_{P}(t)$. The data is based on $10^{4}$ realizations on a system
    of $L=256$, and the maximum observed time was $\approx 180L^{2}$.}
\end{figure}

We define the quench time of the system $T_{q}$ as the time taken for a given
realization to reach its final state. If one considers the distribution of this
quantity, denoted as $P(t)$, on a substantial enough system size, distinct
features associated with each of the topological behaviors studied are apparent
(see Fig.\ref{fig:time_distributions}~(a)).

The times associated with realizations that reach either a ground or on-axis
stripe state are easily visible in Fig.~\ref{fig:time_distributions}~(a). The
left and right hand peaks are associated with cases that reach the ground or
on-axis stripe states respectively. However, the times associated with the slow
decay of the off-axis stripes are in the tail of the distribution, which is
small in magnitude and difficult to resolve on a linear scale. In order to
better illustrate the long--lived nature of these configurations, we consider
the survival probability (see Refs.~\cite{Spirin2001, Olejarz2012}), defined as
\begin{equation}
\mathcal{S}_{P}(t) = 1 - \int_{0}^{t}P(t)\mathrm{d}t.
\end{equation}
The survival probability (shown in Fig.~\ref{fig:time_distributions}~(b)) is
therefore the probability that the system has yet to reach a final state by time
$t$. The long tail of $\mathcal{S}_{P}(t \ge L^{2}) \approx 0.04$ corresponds to
the longer timescale associated with the off-axis winding configurations. At
short times ($t \lessapprox 0.6L^{2}$) the features of
Fig.~\ref{fig:time_distributions}~(a) are also visible.

\subsection{Energy progression}

\begin{figure}[h!]
  \includegraphics[width=\columnwidth]{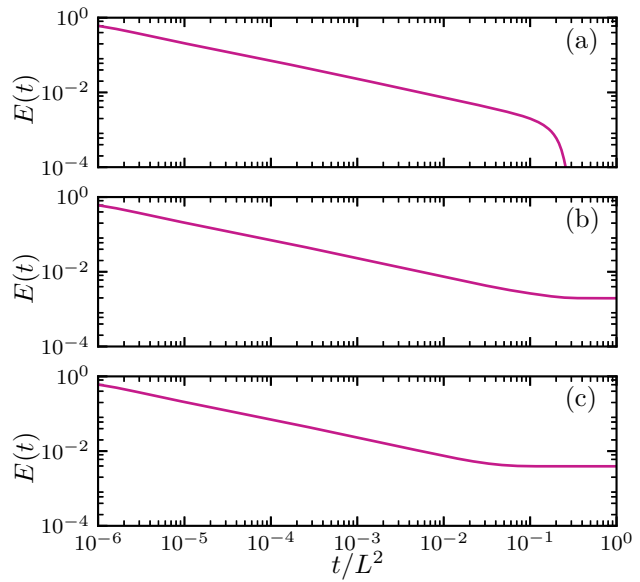}
  \caption{(Color online) time evolution of the mean energy (normalized) for
    cases that evolve to (a) a ground state, (b) an on--axis striped state and
    quenches that (c) evolve through an off--axis winding configuration.}
	\label{fig:energy_progressions}
\end{figure}

The energy progression for each of the three behaviors is given by
Fig.~\ref{fig:energy_progressions}. As one should expect, the energy decays as a
power law until the system is close to reaching its final state, where it
rapidly drops to its final value. In the case of the stripe behaviors, the
energy progressions exhibit a power law decay before plateauing, signifying
maturation of the stripes. At this time the domain interfaces have essentially
no curvature as the system is either in a metastable final state or a
long--lived off-axis configuration, therefore the energy remains constant.

The energy of any inhomogeneous configuration is controlled by the total length
of the interface between the domains, therefore a ``perfect'' off--axis stripe
phase that has an interface rotated $\pm\;\pi / 2$ relative to the lattice axes
is exactly equivalent in energy to an on--axis stripe. One should therefore find
identical final expectation values of the energy in
Fig.~\ref{fig:energy_progressions} (b) \& (c), which is not the case. This can
be explained by considering energy contributions from ``imperfect'' (curved)
interfaces during evolution as well as the presence of rare off--axis
configurations with greater winding numbers. These are off--axis by $>|\pm \pi /
2|$, and therefore greater in interfacial length and subsequently energy. The
presence of at least one such realization in our data explains the slight energy
discrepancy in the final energy values in Fig.~\ref{fig:energy_progressions} (b)
\& (c). Configurations with greater winding numbers are sufficiently rare
(occurance probability $\le$ 0.00015) that they play a negligible role in our
presented results~\cite{Olejarz2012}.

\subsection{Correlations}

The expectation value of the correlation function (Eqn.~\ref{eqn:correlation})
is unity (zero) for ferromagnetically ordered (disordered) configurations. At
early times, short range correlations are established and as the system evolves
long range correlations emerge, signifying the presence of large domain
structures.

\begin{figure*}
  \centering \includegraphics[width=\textwidth]{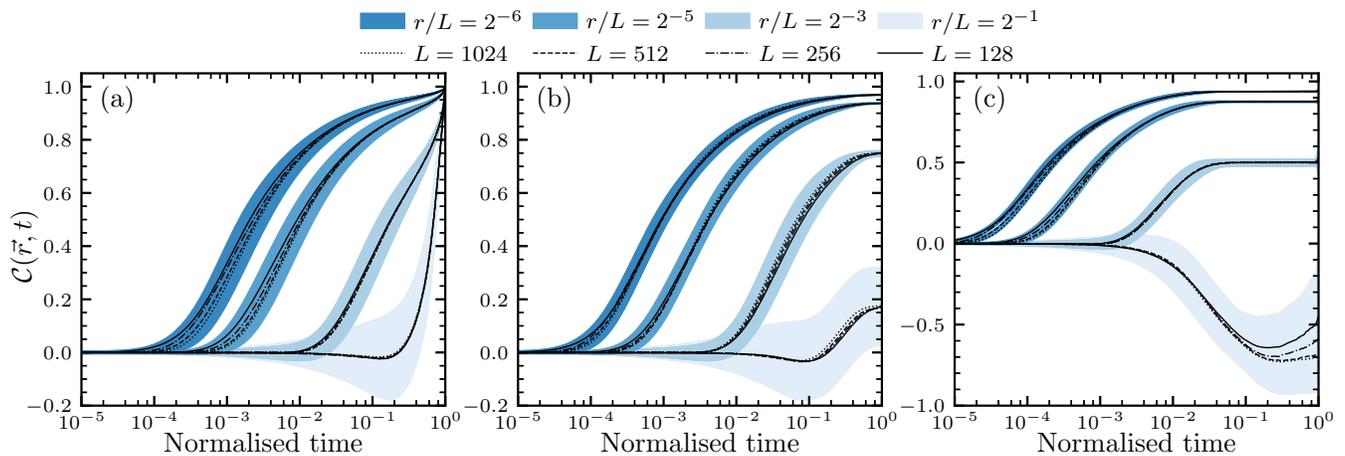}
  \caption{\label{fig:all_correlations} (Color online) Spin product correlation
    functions (black lines) versus time (semi-logarithmic scale) at equivalent
    fractional distances $r / L$ (color coded) on lattices of length $L$ for
    realizations that reach (a) a ground state, (b) an on--axis stripe state or
    evolve through (c) an off--axis configuration. The off--axis stripe
    configurations are terminated early at $t=2L^{2}$. The shaded regions show
    one standard deviation of the correlation over the ensemble. In all cases
    the correlations show good data collapse for all lattice sizes considered.}
\end{figure*}

The progression of the correlation functions for each of the three topologically
distinct behaviours we consider are given by
Fig.~\ref{fig:all_correlations}. The data is presented in normalized time on a
logarithmic scale, encompassing the initial stage where percolating domains
grow, through coarsening to a stable or metastable final state, or in the case
of the long--lived configurations an early termination at time $t=2L^{2}$.

As the system transitions from disordered to homogeneously ordered
(Fig.~\ref{fig:all_correlations}~(a)) the correlations progress from zero to
unity. For on--axis stripe realizations (Fig.~\ref{fig:all_correlations}~(b))
the behavior of the correlation function is similar to that of the ground state
case, however the final expectation does not reach unity as there is always a
remaining stripe of the minority phase. The final value of the correlations in
the metastable final state case depends upon the distance at which they are
measured and the distribution of the stripe widths. Long range correlations are
more likely to involve spins on either side of a phase boundary and therefore
reduce the expectation value.

\begin{figure*}
  \includegraphics[width=\textwidth]{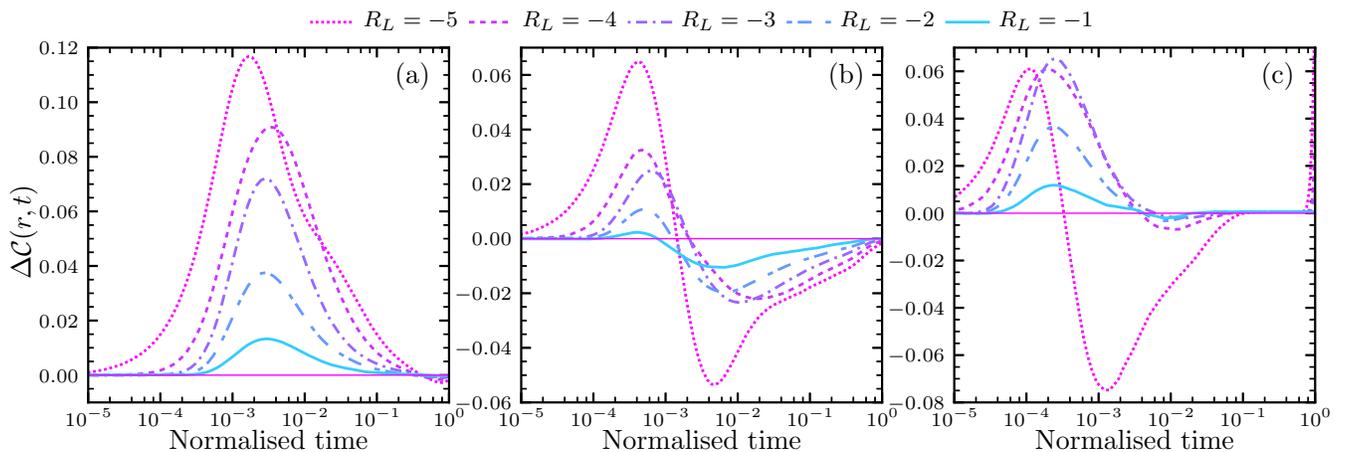}
  \caption{\label{fig:equivalent_differences} (Color online) Differences in
    equivalent correlations $\Delta\mathcal{C}(r, t)$ from lattice sizes of $L$
    and $L^{\prime}$ for both (a) stable and (b) metastable final state
    quenches. The correlations were taken at distances of $r = 2^{-5}L$. The
    ratio between the lattice sizes is expressed as $R_{L} = \log_{2}(L /
    L^{\prime})$. In each case $L^{\prime}=1024$, and $L$ varies over $ 2^{5}
    \le 2^{n} \le 2^{9}$. For all time the maximum differences tend towards zero
    with increasing system length. The solid horizontal line indicates
    $\Delta\mathcal{C}(r, t) =0$.}
\end{figure*}

In the case of the off--axis stripe configurations
(Fig:~\ref{fig:all_correlations}~(c)), the behavior of the correlation function
differs significantly. When the off-axis winding domains have formed, spins
separated by half of the system length are anti-correlated, i.e.\ this distance
spans across the phase boundaries in such realizations.

Fig.~\ref{fig:all_correlations} demonstrates that at fixed fractions of the
system size, the same evolution occurs for systems of any size when compared in
normalized time. This behavior holds from the random initial condition across
the entire coarsening process to the final state, and is distinct for each of
the three topological behaviours.

At normalized times of around $10^{-3} \le t \le 10^{-2}$, the agreement in the
correlations is poorest, particularly for the smallest system sizes
considered. As time progresses or system size increases, the correlations
quickly collapse onto universal curves characteristic of the correlation at
specified fractional distances. We investigate the disagreement in correlations
obtained at the same fractional distance $r / L$ on different systems sizes by
computing
\begin{equation}
  \Delta\mathcal{C}(r, t) = \mathcal{C}(r, t) - \mathcal{C}^{\prime}(r, t).
\end{equation}
$\mathcal{C}^{\prime}$ is the correlation obtained on a system of length
$L^{\prime} = bL$ at a distance of $br$, and $\mathcal{C}$ is the correlation
from a system of length $L$ taken at a distance of $r$. The scale factor $b$ is
always an integer power of $2$. Examples of the difference in equivalent
correlations are given by Fig~\ref{fig:equivalent_differences}. As the lattice
sizes increases, the mismatch in correlations decays to a negligible
degree. Therefore, away from small system sizes, the behaviour of small systems
parallels that of arbitrarily larger cases.

\begin{figure*}[ht!]
  \includegraphics[width=\textwidth]{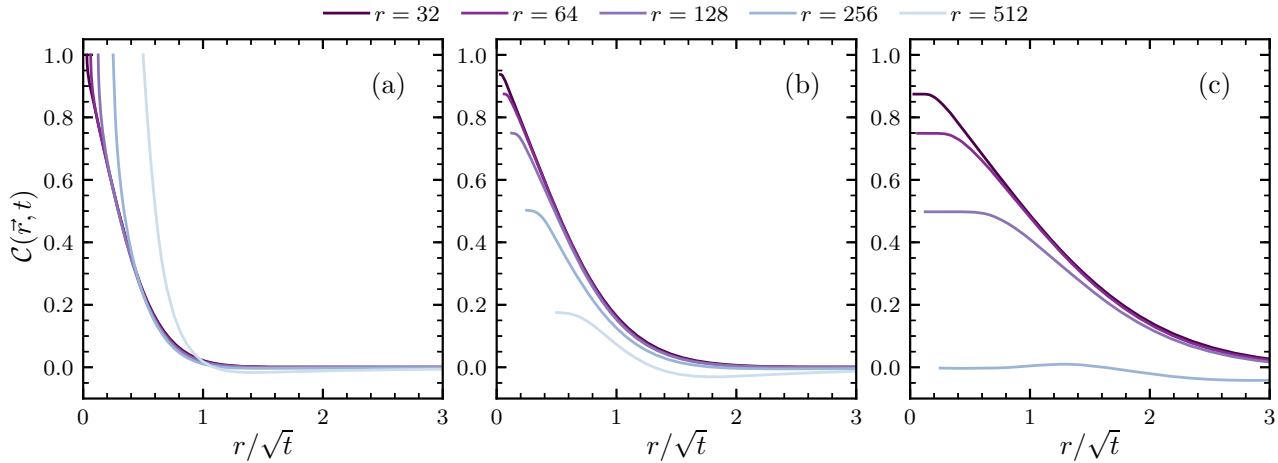}
  \caption{\label{fig:dynamical_scaling} Correlations as a function of
    $r/\sqrt{t}$ for (a) stable and (b) metastable final state quenches. The
    data is averaged over only the active simulations at each time. In each case
    one can see the typical data collapse associated with dynamical scaling, and
    also the failure of this collapse when the system has left the scaling
    regime.}
\end{figure*}


\section{Discussion and concluding remarks}

The dynamic scaling hypothesis asserts that the evolution of the system is
governed by the growth of a characteristic length $R(t) \propto t^{1/z}$ (where
$z=2$ for the 2D Ising model)~\cite{Bray1991, Bray1993, Hohenberg1997}. In the
regime where this length is greater than the lattice spacing, but much smaller
than the system size, the two-point correlation functions collapse onto a
universal curve of the form~\cite{Bray1991, Bray1993}
\begin{equation} \mathcal{C}(r, t)
\propto G \left( \frac{r}{R(t)} \right) \propto G \left( \frac{r}{t^{1/z}}
\right). \label{eqn:scaling_law}
\end{equation}
The characteristic length associated with the coarsening process, $R(T)$, is a
measure of the typical domain size. In the case of the 2D kinetic Ising model
the exponent is $z=2$~\cite{Bray1993}. Spins at shorter distances rapidly
correlate and spins at distances greater than the characteristic length are
uncorrelated. This pattern extends to longer scales as the system coarsens and
domains grow.

When the system is no longer in the regime where $R(t) \ll L$, it has left the
dynamical scaling regime and the growth of $R(t)$ is inhibited by boundary
effects. Even at times where long range correlations are low, the spanning
domains that determine the final state of the system have emerged, saturating
the domain growth. Outwith this regime the functional form of data collapse
typically associated with the dynamical scaling hypothesis fails (see
Fig.~\ref{fig:dynamical_scaling}). This is presumably a manifestation of similar
differences in the universality classes of the bulk and boundary
regions~\cite{Izmailian2010} at criticality.

At such times it is reasonable to assume that the data collapse could be
restored by considering a term to control finite size
effects~\cite{Corberi2017}, i.e.\
\begin{equation}
C(r, t) = G\left( \frac{r}{R(t)}, \frac{L}{R(t)} \right).
\end{equation}
However, the correlations associated with the ground state cases
(Fig.~\ref{fig:dynamical_scaling}~(a)) are shifted to the right, whereas the
cases involving either stable or unstable winding domains
(Fig.~\ref{fig:dynamical_scaling}~(b)--(c)) are shifted to the left. This
suggests that any rescaling to account for finite size effects needs to be
topologically aware.

Here we argue that in finite systems at all scales and times there is data
collapse of the expectation of the correlation functions. This applies for each
of the topologically distinct behaviors studied and holds over the full
evolution of the system, including the classical dynamical scaling regime.

In conclusion, we have shown that for finite square lattice Ising ferromagnets
evolving under Glauber dynamics, there is an equivalence in how correlations
evolve at fractional distances within the system in normalized time. Not only
does this equivalence hold in the regime where the evolution is well described
by the dynamical scaling hypothesis, but also after the system has left this
scaling regime. It would be of interest to see the exploration of this behavior
in other systems where the exponent $z$ is the same as the 2D Ising model.
Furthermore, there is also scope for investigation in other systems displaying
curvature driven coarsening described by different forms of this dynamical
scaling law, where the exponent $z$ can vary from system to system, as well as
in different regimes of the phase ordering process of a particular
system~\cite{Camley2011}.

J.D.\ acknowledges support from EPSRC DTA grant EP/N509760/1. We acknowledge the
ARCHIE-WeSt High Performance Computer (www.archie-west.ac.uk) based at the
University of Strathclyde, as well as EPSRC grant EP/P015719/1. The authors
would like to thank Oliver Henrich and Sidney Redner for helpful comments in
preparation of this manuscript.


\bibliography{references.bib}

\end{document}